\documentclass[apj,numberedappendix]{emulateapj}

\usepackage{apjfonts}
\shorttitle{ALMA Observations of HL~Tau}
\shortauthors{The ALMA Partnership et al.}

\slugcomment{Submitted to ApJL on 25 February 2015;  accepted on 14 March 2015}

\newcommand{\HLt}{HL~Tau}

\newcommand\kms{km~s$^{-1}$}

\newcommand\mjb{mJy~beam$^{-1}$}
\newcommand\mujb{$\mu$Jy~beam$^{-1}$}

\newcommand\h{^{\rm h}}
\newcommand\m{^{\rm m}}
\newcommand\s{^{\rm s}}
\newcommand{\lsun}{L$_\odot$}
\newcommand{\msun}{M$_\odot$}
\newcommand{\Lk}{LkH$\alpha$358}
\newcommand{\co}{$^{12}$CO~(1-0)}
\newcommand{\cotwoone}{$^{12}$CO~(2-1)}
\newcommand{\hcop}{HCO$^+$~(1-0)}
\newcommand{\hcn}{HCN~(1-0)}
\newcommand{\cn}{CN~(1-0)}
\newcommand{\ha}{H$\alpha$}
\newcommand{\sulf}{[SII]}

\begin{document}

\title{First Results from High Angular Resolution ALMA Observations
  Toward the HL~Tau Region}

\author{ALMA Partnership,
C. L. Brogan\altaffilmark{1},
L. M. P\'erez\altaffilmark{2,$*$},
T. R. Hunter\altaffilmark{1},
W. R. F. Dent\altaffilmark{3,4},
A. S. Hales\altaffilmark{3,1},
R. E. Hills\altaffilmark{5},
S. Corder\altaffilmark{3,1},
E. B. Fomalont\altaffilmark{3,1},
C. Vlahakis\altaffilmark{3,4},
Y. Asaki \altaffilmark{6,7},
D. Barkats\altaffilmark{3,4},  
A. Hirota\altaffilmark{3,6},  
J. A. Hodge\altaffilmark{1,$*$},
C. M. V. Impellizzeri\altaffilmark{3,1},
R. Kneissl\altaffilmark{3,4},
E. Liuzzo\altaffilmark{8},
R. Lucas\altaffilmark{9},
N. Marcelino\altaffilmark{8},
S. Matsushita\altaffilmark{10},
K. Nakanishi\altaffilmark{3,6},  
N. Phillips\altaffilmark{3,4}, 
A. M. S. Richards\altaffilmark{11},
I. Toledo\altaffilmark{3},
R. Aladro\altaffilmark{4},
D. Broguiere\altaffilmark{12}, 
J. R. Cortes\altaffilmark{3,1},
P. C. Cortes\altaffilmark{3,1},
%V. Dhawan\altaffilmark{2},
D. Espada\altaffilmark{3,6},
F. Galarza\altaffilmark{3},
D. Garcia-Appadoo\altaffilmark{3,4}, 
L. Guzman-Ramirez\altaffilmark{4},  
E. M. Humphreys\altaffilmark{13},  
T. Jung\altaffilmark{14},   
S. Kameno\altaffilmark{3,6}, 
R. A. Laing\altaffilmark{13},     
S. Leon\altaffilmark{3,4},
G. Marconi\altaffilmark{3,4}, 
A. Mignano\altaffilmark{8},
B. Nikolic\altaffilmark{5},
L. -A. Nyman\altaffilmark{3,4}, 
M. Radiszcz\altaffilmark{3}, 
A. Remijan\altaffilmark{3,1},
J. A. Rod\'on\altaffilmark{4},  
T. Sawada\altaffilmark{3,6},
S. Takahashi\altaffilmark{3,6},
R. P. J. Tilanus\altaffilmark{15},   
B. Vila Vilaro\altaffilmark{3,4}, 
L. C. Watson\altaffilmark{4},
T. Wiklind\altaffilmark{3,4},
E. Akiyama\altaffilmark{6},
E. Chapillon\altaffilmark{12,16,18},
I. de Gregorio-Monsalvo\altaffilmark{3,4},
J. Di Francesco\altaffilmark{17},
F. Gueth\altaffilmark{12},
A. Kawamura\altaffilmark{6},
C.-F. Lee\altaffilmark{10},
Q. Nguyen Luong\altaffilmark{6}, 
J. Mangum\altaffilmark{1},
V. Pietu\altaffilmark{12},
P. Sanhueza\altaffilmark{6},
K. Saigo\altaffilmark{6},
S. Takakuwa\altaffilmark{10}, 
C. Ubach\altaffilmark{1},
T. van Kempen\altaffilmark{15}, 
A. Wootten\altaffilmark{1},
A. Castro-Carrizo\altaffilmark{12},
H. Francke\altaffilmark{3},
J. Gallardo\altaffilmark{3},
J. Garcia\altaffilmark{3},
S. Gonzalez\altaffilmark{3}, 
T. Hill\altaffilmark{3,4},
T. Kaminski\altaffilmark{4}, 
Y. Kurono\altaffilmark{3, 6},
H.-Y. Liu\altaffilmark{10},
C. Lopez\altaffilmark{3},
F. Morales\altaffilmark{3},
K. Plarre\altaffilmark{3},
G. Schieven\altaffilmark{17},  
L. Testi\altaffilmark{13},
L. Videla\altaffilmark{3},
E. Villard\altaffilmark{3,4},
P. Andreani\altaffilmark{13},
J. E. Hibbard\altaffilmark{1},
K. Tatematsu\altaffilmark{6}
}

% NRAO Cville
\altaffiltext{1}
{National Radio Astronomy Observatory, 520 Edgemont Rd, Charlottesville, VA, 22903, USA}

% NRAO Socorro
\altaffiltext{2}
{National Radio Astronomy Observatory, P.O. Box O, Socorro, NM 87801, USA}

% JAO
\altaffiltext{3}
{Joint ALMA Observatory, Alonso de C\'ordova 3107, Vitacura, Santiago, Chile}

% ESO Chile
\altaffiltext{4}
{European Southern Observatory, Alonso de C\'ordova 3107, Vitacura, Santiago, Chile}

% Cambridge
\altaffiltext{5}
{Astrophysics Group, Cavendish Laboratory, JJ Thomson Avenue, Cambridge, CB3 0HE, UK}

% NAOJ
\altaffiltext{6}
{National Astronomical Observatory of Japan, 2-21-1 Osawa, Mitaka, Tokyo 181-8588, Japan}

% JAXA/ISAS
\altaffiltext{7}
{Institute of Space and Astronautical Science (ISAS), Japan Aerospace Exploration Agency (JAXA), 3-1-1 Yoshinodai, Chuo-ku, Sagamihara, Kanagawa 252-5210 Japan}

% Bologna
\altaffiltext{8}
{INAF, Istituto di Radioastronomia, via P. Gobetti 101, 40129 Bologna, Italy}

% Robert Lucas
\altaffiltext{9}
{Institut de Plan\'etologie et d'Astrophysique de Grenoble (UMR 5274), BP 53, 38041, Grenoble Cedex 9, France}
	
% ASIAA
\altaffiltext{10}
{Institute of Astronomy and Astrophysics, Academia Sinica, P.O. Box 23-141, Taipei 106, Taiwan}

% Manchester
\altaffiltext{11}
{Jodrell Bank Centre for Astrophysics, School of Physics and Astronomy, University of Manchester, Oxford, Road, Manchester M13 9PL, UK}

% IRAM
\altaffiltext{12}
{IRAM, 300 rue de la piscine 38400 St Martin d'H\`eres, France}

% ESO Garching
\altaffiltext{13}
{European Southern Observatory, Karl-Schwarzschild-Str. 2, D-85748 Garching bei M\"unchen, Germany}

% KASI
\altaffiltext{14}{
Korea Astronomy and Space Science Institute, Daedeokdae-ro 776, Yuseong-gu, Daejeon 305-349, Korea}

% Leiden
\altaffiltext{15}
{Leiden Observatory, Leiden University, P.O. Box 9513, 2300 RA Leiden, The Netherlands}

\altaffiltext{16}
{Univ. Bordeaux, LAB, UMR 5804, 33270 Floirac, France}

% NRC NAASC
\altaffiltext{17}
{National Research Council Herzberg Astronomy \& Astrophysics, 5071
  West Saanich Road, Victoria, BC V9E 2E7, Canada}

\altaffiltext{18}
{CNRS, LAB, UMR 5804, 33270 Floirac, France}

\altaffiltext{$*$}
{NRAO Jansky Fellow}

\begin{abstract}

\noindent 

We present Atacama Large Millimeter/submillimeter Array (ALMA) observations from the 
2014 Long Baseline Campaign in dust continuum and spectral line 
emission from the \HLt\ region. The continuum images at wavelengths
of 2.9, 1.3, and 0.87~mm have unprecedented angular resolutions of
0\farcs075 (10~AU) to 0\farcs025 (3.5~AU), revealing an astonishing level of
detail in the circumstellar disk surrounding the young solar analogue 
\HLt, with a pattern of bright and dark rings observed at all wavelengths. 
By fitting ellipses to the 
most distinct rings, we measure precise values for the disk inclination 
($46.72\arcdeg\pm0.05\arcdeg$) and position angle ($+138.02\arcdeg\pm0.07\arcdeg$). 
We obtain a high-fidelity image of the 1.0~mm spectral index ($\alpha$), which ranges from $\alpha\sim2.0$ in the optically-thick central peak and two brightest rings, increasing to 2.3-3.0 in the dark rings. The dark rings are not devoid of emission, and we estimate a grain emissivity index of 0.8 for the innermost dark ring and lower for subsequent dark rings, consistent with some degree of grain growth and evolution.  Additional clues that the rings arise from planet formation include 
an increase in their central offsets with radius and the presence of numerous orbital resonances. At a resolution of 35~AU, we resolve the molecular component of the disk in \hcop\/ 
which exhibits a pattern over LSR velocities from 2-12 \kms\/ 
consistent with Keplerian motion around a $\sim$1.3\msun\/ star, although complicated by absorption at low blue-shifted velocities. We also serendipitously detect and resolve the nearby protostars XZ~Tau (A/B) and \Lk\/ at 2.9~mm.
\end{abstract}


\section{Introduction}

Inside the Taurus star-forming complex, \HLt\/ is a young star located in a molecular ridge of length $\sim 0.05$~pc, which forms part of the wall of a large-scale bubble seen in both $^{13}$CO and faint scattered light \citep{Welch2000,Anglada2007}. This material contributes to the estimated $\sim$24-33 magnitudes of visual extinction ($A_{\rm V}$) towards \HLt\/ \citep{Close1997,Menshchikov1999}.   Three other less obscured protostars lie toward the edges of this molecular ridge, in order of separation from \HLt\/: XZ~Tau, \Lk\/, and HH30 \citep[e.g.,][]{Moriarty2006}. Together these four protostars make up the ``\HLt\ region'' in L1551. In this paper we adopt the standard mean distance to Taurus of 140~pc \citep{Rebull04} for the \HLt\ region\footnote{However, Very Long Baseline Interferometry measurements indicate that the line-of-sight depth of the Taurus complex is $\sim 20$~pc, so that reported linear distances could be in error by up to $15\%$ \citep[e.g.,][]{Loinard2013}}.

As a result of the high extinction, the central star of \HLt\/ has not been directly detected at optical wavelengths, and only a conical reflection nebula has been observed \citep[e.g.][]{Stapelfeldt1995}. Near-infrared (NIR) images reveal a point-like object that has been attributed to direct stellar radiation; though, scattered emission from a central hot disk is also a likely explanation \citep{Close1997,Menshchikov1999}. Based on high dispersion optical spectroscopy, \HLt\/ is classified as spectral type K5$\pm 1$ \citep{White2004}. Analyses of the full SED find bolometric luminosities ranging from 3.5 to 15~\lsun\/, and classify it as a Class I-II protostar with evidence of an extended envelope and a circumstellar disk \citep{Kenyon1995,Menshchikov1999,Robitaille2007}. In all, the stellar properties of \HLt\/ still have a high degree of uncertainty despite extensive study.

Measurements in the millimeter regime suffer far less from extinction effects, and the \HLt\/ disk is one of the brightest at these wavelengths \citep{Andrews2005}; hence \HLt\ has been a favored interferometric target over the last two decades \citep[][to name a few]{Sargent1991,Mundy1996,Lay1997,Kitamura2002,Looney2000,Guilloteau2011,Stephens2014}. 
The highest angular resolution mm observations to date have a $0\farcs13$ (18~AU) beam at 1.3~mm in which the disk is resolved with an outer radius of 120~AU at position angle +136$\arcdeg$ and an inclination (from face-on) of 40$\arcdeg$ \citep{Kwon2011}. Modeling of previous millimeter data suggest an HL~Tau disk mass of $M_d\sim 0.03-0.14$~M$_{\odot}$ \citep[][]{Robitaille2007,Guilloteau2011,Kwon2011}. This high disk mass is 
within an order of magnitude of previous HL~Tau stellar mass estimates ranging from 0.55 to 1.2 ~M$_{\odot}$ \citep[]{Sargent1991,Close1997,White2004}, suggesting that the disk may be close to being gravitationally unstable.  

At longer (cm) wavelengths,  free-free emission from an outflow jet orthogonal to the disk plane contributes to the continuum \citep{Wilner1996, Rodmann2006,Greaves2008,Carrasco2009}. Also known as HH150-151, this jet is seen in optical and near-IR shock-tracing lines such as [SII] and [FeII], extending up to $\sim 10,000$~AU from the star at velocities of 100-200 \kms\/ \citep{Mundt1990,Krist2008}. The inner jet has been traced down to $\sim$100~AU from the disk \citep{Pyo2006,Takami2007,Beck2010}.  At mm wavelengths, CO emission traces a lower-velocity, entrained component of this outflow,  mainly showing red-shifted gas on the southwest side of the disk \citep{Cabrit1996,Monin1996}.  The central few hundred AU of this outflow was also observed using the SMA in the $^{12}$CO (3-2) line \citep{Lumbreras2014}, and showed a similar dominance of red-shifted gas.  Detailed study of the molecular gas associated with the disk 
has heretofore been plagued by spatial and kinematic confusion due to the extended ridge and outflow emission near the ambient velocity of v$_{lsr}=6-7$~\kms\/ \citep[e.g.][]{Cabrit1996}.

Overall, \HLt\/ is an excellent example of a system just emerging from its protostellar cocoon (i.e. evolving from SED Class I to Class II), which contains a massive compact disk as well as highly collimated
outflow. The system is young \citep[$\leq$1-2~Myr based on the cluster age of Taurus,][]{Briceno2002}, and the high disk mass makes it an ideal candidate for disk instability and early planet formation \citep[e.g.][]{Nero2009}. In this paper, we present multi-wavelength ALMA Science Verification (SV) continuum and spectral line observations from the 2014 Long Baseline Campaign (LBC) of the \HLt\ region. 
These data dramatically demonstrate the revolutionary impact that the full sensitivity and resolution of ALMA will have on the field of star and planet formation. In this work, we aim to give an initial taste of the incredible richness extant in these first ALMA long baseline observations of a protoplanetary disk, and we expect that in time the community will uncover the full scope of what is possible with these data. 

\section{Observations and Data Reduction}

The ALMA SV data on \HLt\ was taken between 
2014 October 14 and November 14 with baselines ranging from 15~m to 15.2~km. \citet{ALMA15a} gives the detailed tuning frequencies for each band. To summarize, the 1.3~mm and 0.87~mm data (Bands 6 and 7) have four 2~GHz (1.875~GHz usable) spectral windows (spws) with 128 channels per spw (hereafter referred to as a ``continuum'' spw). The 2.9~mm data (Band 3) were taken with two different tunings. One used three continuum spws, along with two narrow bandwidth spws centered on \hcop\/ and HCN (1-0), each with a channel resolution of $\sim 0.21$~\kms\/ and a velocity resolution of 0.42 \kms\/. The other 2.9~mm tuning used two continuum spws, one narrow bandwidth window on \co\/, and another spanning four hyperfine transitions of CN (1-0). These data have a channel resolution of $\sim 0.16$~\kms\/ and a velocity resolution of 0.32 \kms\/. The scheduling blocks (SBs) were designed to run for 60-70 minutes, with approximately 30 minutes on-source per execution, and each included about 30 antennas. Each of the 2.9~mm SBs were executed seven times, while the 1.3~mm and 0.87~mm SBs were executed nine and ten times respectively. The start times were staggered to obtain good hour angle coverage for each band, and hence good uv-coverage. It cannot be overstated how crucial the uv-coverage is to the successful imaging of the complex structure of \HLt\/.

The calibration and imaging for the continuum and spectral line data are described in \S~\ref{Details}. The final synthesized beams and rms noise levels for the continuum data are 
given in Table~\ref{Cont}. Cubes with 0.25 \kms\/ velocity width channels were made for the four observed lines at $1\farcs1$ angular resolution. Additionally, an \hcop\/ cube with $0\farcs25$ resolution was also created (rms noise levels for the cubes are given in \S~\ref{Details}). All of the images described in this paper are publicly available from the ALMA
Science Verification page\footnote{\label{fnote}The raw data,
  calibration scripts (including detailed absolute flux scale
  information), and images are available from
  \url{http://almascience.org/alma-data/science-verification}. Additional details
  along with the full imaging scripts are available from
  \url{http://casaguides.nrao.edu/index.php?title=ALMA2014\_LBC\_SVDATA}.}.

\begin{figure*}
\centering
\includegraphics[width=0.7\textwidth]{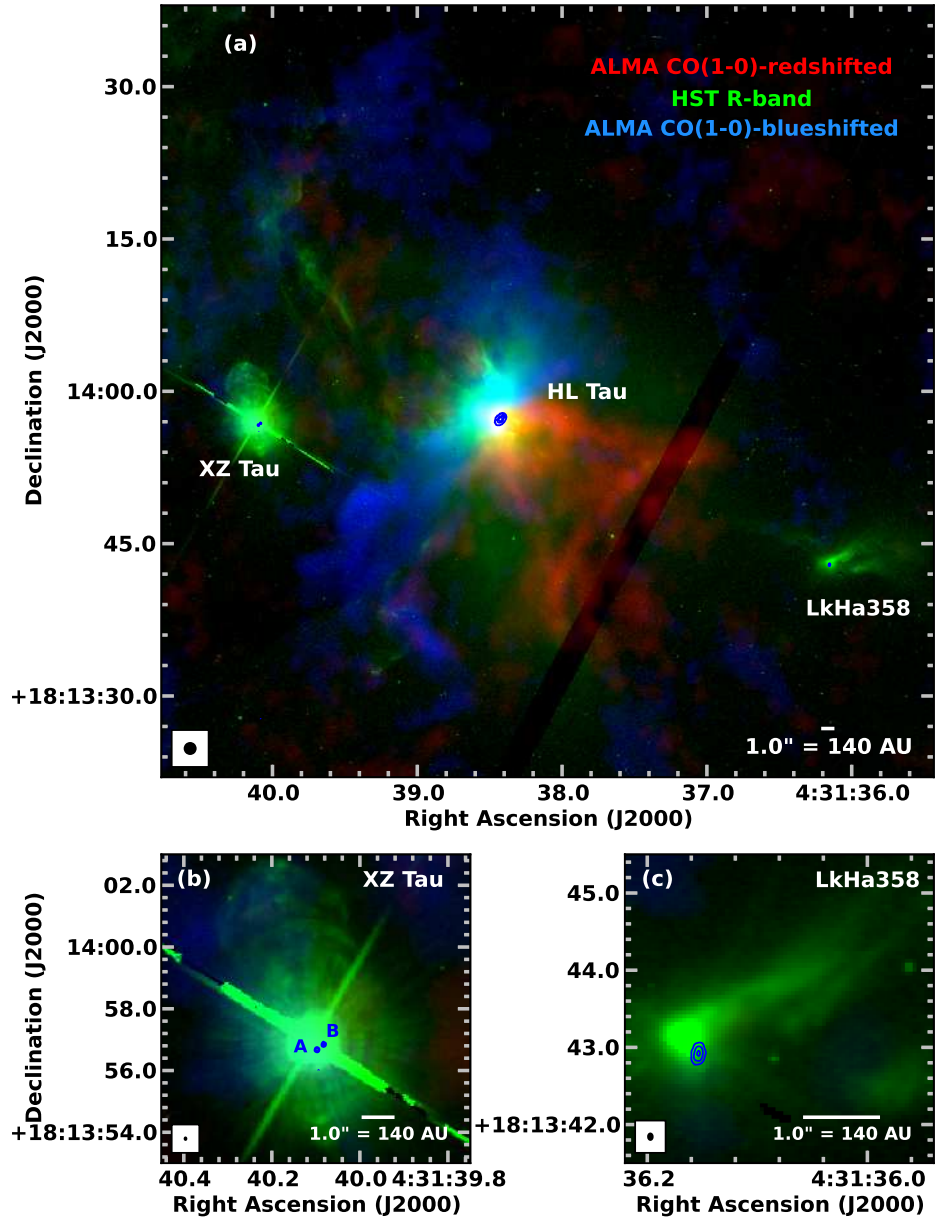}
\caption{Panel (a) shows an overview of the \HLt\ region with red and blue colors mapped to the integrated intensity of the  redshifted (7.25 to 20 \kms\/) and blueshifted (0.0 to 6.0 \kms\/) \co\/, respectively. The ALMA $1\farcs1$ resolution \co\/ images (synthesized beam shown in the lower left) have not been corrected for primary beam attenuation; the displayed field of view corresponds to the $\sim 15\%$ power point.  Green shows an {\em HST} R-band image (this band includes stellar continuum, \ha\/, and \sulf\/).  The darker diagonal stripe across the image (visible in some viewers) corresponds to a gap in the {\em HST} ACS WFC detector. Panels (b) and (c) show zoomed in views of XZ~Tau (A and B) and \Lk\/, respectively. All three panels show ALMA 2.9~mm primary beam corrected continuum contours overlaid in blue at 33$\times$(4, 14,19) \mujb\/ (the corresponding synthesized beam is shown in the lower left of panels (b) and (c), also see Table~\ref{Cont}). The {\em HST} image has been precessed to epoch 2014.83 using the proper motion derived in \S3.1.1.}
\label{CO}
\end{figure*}

\section{Results}

\begin{deluxetable*}{llccccc}[t]
\tabletypesize{\footnotesize}
\tablewidth{0pc}
\tablecaption{Measured Continuum Parameters\tablenotemark{a}}
\tablecolumns{7}
\tablehead{
\colhead{Source} & \multicolumn{2}{c}{Position\tablenotemark{b}} & \colhead{rms} & \colhead{Peak\tablenotemark{c}} & \colhead{Flux Density\tablenotemark{c}} & \colhead{Size\tablenotemark{c}}\\   
\colhead{} & \colhead{R.A. (J2000)} & \colhead{Dec. (J2000)} & \colhead{(\mujb)} & \colhead{(\mjb)} & \colhead{(mJy)} & \colhead{(mas x mas [deg])}
}
\startdata
\cutinhead{2.9 mm (Band 3, 101.9 GHz) 85.3 $\times$ 61.1 mas, PA$ = -179\arcdeg$}
HL Tau    & 04:31:38.4253  (0.0002)  & 18:13:57.240 (0.004)   &  9 &  4.13 (0.03) & 74.3 (0.4)  & $\sim 1750$ \\
XZ Tau A  & 04:31:40.09714 (0.00004) & 18:13:56.674 (0.001)   & 24 &  1.99 (0.04) & 2.71 (0.08) & $<$ 54 \\
XZ Tau B  & 04:31:40.08218 (0.00007) & 18:13:56.845 (0.001)   & 24 &  1.30 (0.04) & 1.83 (0.08) & $<$ 54 \\
\Lk       & 04:31:36.15383 (0.00007) & 18:13:42:919 (0.002)   & 26 &  1.07 (0.03) & 3.7 (0.1) & 150 x 84 [+170] (6x3 [3])\\
\cutinhead{1.3 mm (Band 6, 233.0 GHz) 35.1 $\times$ 21.8 mas, PA$ = +11\arcdeg$}
HL Tau    & 04:31:38.42548 (0.00008) & 18:13:57.242 (0.001) & 11 &  6.48 (0.03) &  744.1 (1.5) & $\sim 1960$ \\ 
\cutinhead{0.87 mm (Band 7, 343.5 GHz) 29.9 $\times$ 19.0 mas, PA$ = -176\arcdeg$}
HL Tau    & 04:31:38.4254 (0.0001) & 18:13:57.242 (0.001)   & 23 & 11.56 (0.07) & 2140.8 (3.7) & $\sim 1960$  \\
\cutinhead{1.0 mm (Combined Band 6+7, 287.2 GHz) 33.5 $\times$ 21.1 mas, PA$ = +9\arcdeg$}
HL Tau    & 04:31:38.42545 (0.00009) & 18:13:57.242 (0.001) & 12 &  9.79 (0.04) & 1441.5 (1.8) & $\sim 1960$ 
\enddata
\tablenotetext{a} {For each band, the mean continuum frequency and synthesized beam is given after the wavelength. Uncertainties are given in parenthesis after each quantity. All measurements are made from images corrected for primary beam attenuation.}
\tablenotetext{b} {The peak positions (and uncertainties) were measured using Gaussian
  fitting, for \HLt\ the fit was restricted to the region inside the smallest radii gap.}
\tablenotetext{c}{For XZ~Tau and \Lk\/, the peak intensity, flux density and size (or upper limit) were measured using Gaussian fitting. For \HLt\, the peak intensity corresponds to the peak pixel value with an uncertainty of $3\sigma$ (where $\sigma$ is given in the rms column). The flux density was measured within the 4$\sigma$ contour, and the uncertainty is [No. Independent Beams]$^{0.5}\times 3\sigma$. Uncertainties for these quantities do not include the $5\%$ absolute flux uncertainty. The size of \HLt\ was estimated from the length of the line passing through the continuum peak at a position angle of $+138\arcdeg$ and where it crosses the 4$\sigma$ contour on each side of the disk.}
\label{Cont}
\end{deluxetable*}

\begin{figure*}
\centering
\includegraphics[width=0.8\textwidth]{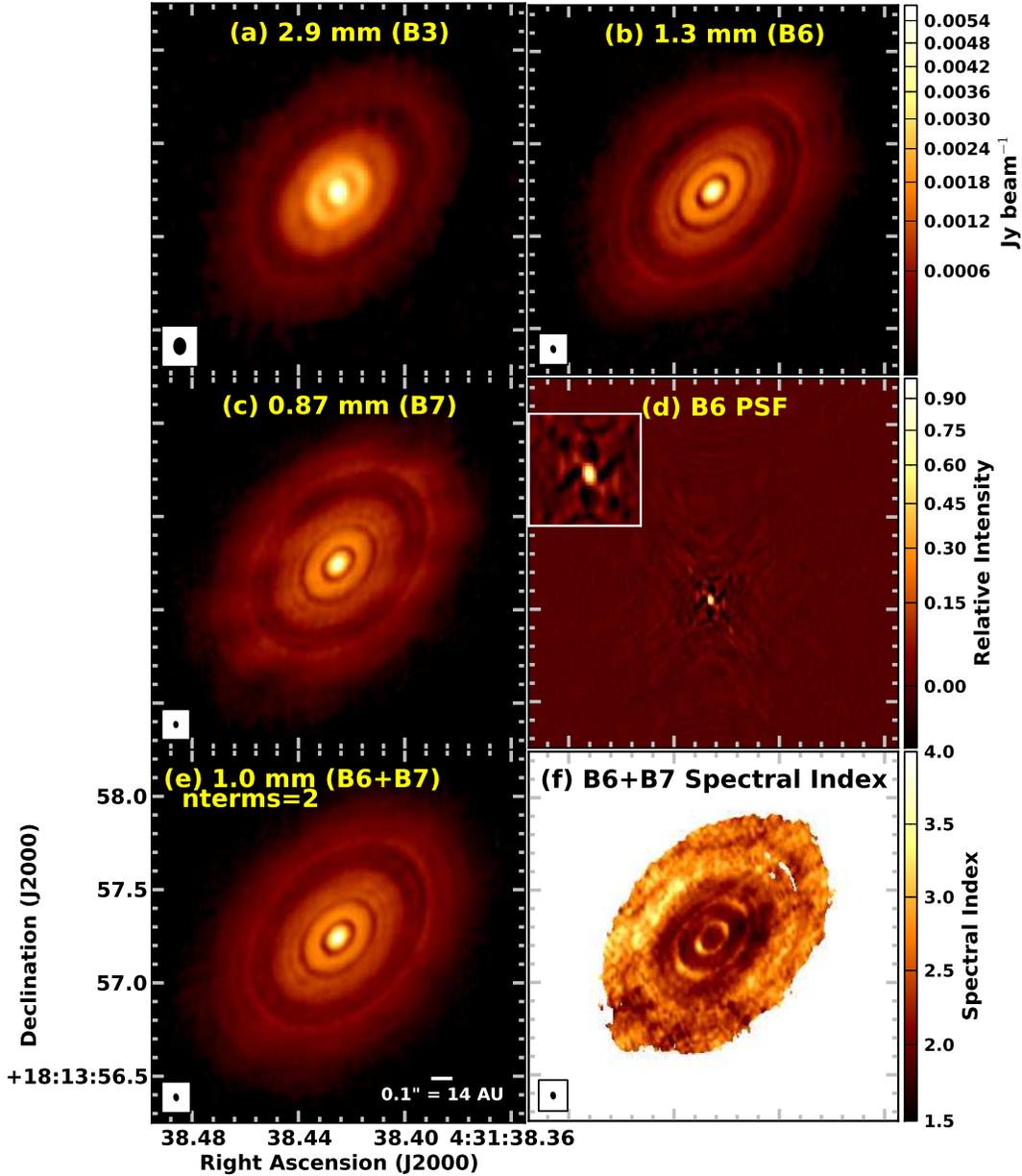}
\caption{Panels (a), (b), and (c) show 2.9, 1.3, and 0.87~mm ALMA continuum images of \HLt\/. Panel (d) shows the 1.3~mm psf for the same FOV as the other panels, as well as an inset with an enlarged view of the inner 300~mas centered on the psf's peak (the other bands show similar patterns). Panels (e) and (f) show the image and spectral index maps resulting from the combination of the 1.3 and 0.87~mm data. The spectral index ($\alpha$) map has been masked where $\alpha/\alpha_{error}<4$.
The synthesized beams are shown in the lower left of each panel, also see Table~\ref{Cont}. The range of the colorbar shown for panel (b), at 1.3~mm, corresponds to $-2\times$rms to $0.9\times$ the image peak, using the values in Table~\ref{Cont}. The colorscales for panels (a), (c) and (e) are the same except using the values of rms and image peak corresponding to each respective wavelength in Table~\ref{Cont}.}
\label{hltau}
\end{figure*}

Figure~\ref{CO}a demonstrates the complex outflow emission that fills the \HLt\ region as traced by both atomic and molecular gas. The strong \ha\/ and \sulf\/ emission (observed with the {\em HST}) and weak blue-shifted \co\/ to the NE of \HLt\, as well as the strong red-shifted \co\/ emission to the SW (observed with ALMA) are perpendicular to the disk, consistent with previous observations (see \S1). Interestingly, the \ha\/, \sulf\/, and blue-shifted \co\/ emission to the SE of \HLt\ {\em parallel} to the disk are difficult to reconcile with a simple disk/outflow scenario, suggesting that the blue-shifted outflow has broken out of the parental core \citep{Monin1996}, or that there is another -- as yet unidentified -- driving source. Unfortunately, the \co\/ data are missing significant flux (due to a lack of short spacings), and have insufficient sensitivity in the outer portions of the field of view to warrant deeper analysis of its properties. Figs.~\ref{CO}b, and c show zoomed in views of our serendipitous detections of XZ~Tau (A and B), and \Lk\/; no other continuum sources above the local $4\sigma$ level were detected.

\subsection{HL~Tau}

\subsubsection{Position and Proper Motion}

The fitted position for \HLt\/ in each of the ALMA images is given in Table~\ref{Cont}.  The phase calibrator positions are accurate to $<1$~mas and the positions are consistent between the three observed bands to better than 2~mas \citep[consistent with dedicated LBC astrometry experiments, see][]{ALMA15a}; thus, we assume 2~mas as the absolute ALMA position uncertainty.  The position reported by \citet{Kwon2011} from 1.3~mm CARMA observations is $04\h31\m38\s.418$ $+18\arcdeg13\arcmin57\farcs37$ (J2000, epoch 2009.08). The phase calibrator for CARMA observations (J0510+1800) had a position accurate to better than 1~mas, and we assume an overall astrometric uncertainty of 5~mas for this measurement. Adding the two uncertainties in quadrature, the measured angular separation between the CARMA position and the 1.0~mm ALMA position (epoch 2014.83, Table~\ref{Cont}) is $\Delta$R.A.$=+106.1\pm 5.6$~mas, $\Delta$Dec$=-128.0\pm 5.6$~mas. With a time span of 5.75 years, this separation amounts to a millimeter-derived proper motion of $v_{R.A.}=+18.5\pm 1$ mas year$^{-1}$, and $v_{Dec.}=-22.3\pm 1$ mas year$^{-1}$. This result is in good agreement with the proper motion adopted by \citet{Guilloteau2011} based on 12 years of Plateau de Bure Interferometer (PdBI) measurements (+14,$-$20 mas year$^{-1}$). For comparison, the weighted average of optical and infrared proper motions for \HLt\ listed in Vizier\footnote{ \url{http://vizier.u-strasbg.fr/viz-bin/VizieR}} since 2000 \citep[e.g.][]{Dias2014,Zacharias2013,Roeser2010} is 
$v_{R.A.}=+2.0\pm 2.4$ mas year$^{-1}$, $v_{Dec.}=-20.9\pm 2.4$ mas year$^{-1}$.  Apparently, the 
optical/IR proper motion for \HLt\ is reasonably accurate in Dec., but is significantly underestimated in R.A. This 
discrepancy is not surprising given that the optical and even near-IR emission from \HLt\ is dominated by 
reflection nebulosity rather than stellar light (see for example Fig.~\ref{CO}a).

\subsubsection{Continuum Emission}

Figures~\ref{hltau}a,b, and c show the \HLt\ continuum at all three observed bands. These images reveal for the first time the remarkable morphology of the \HLt\ protoplanetary disk, with a complex pattern of alternating bright and dark rings. With the caveats of lower spatial resolution at 2.9~mm and reduced phase stability and hence image fidelity at 0.87~mm, the agreement between the different wavelengths is excellent. Although the antenna configuration was not ideal (baselines were concentrated at lengths $< 200$~m and $> 500$~m), a reasonably Gaussian synthesized beam was achieved as demonstrated in Fig.~\ref{hltau}d which shows the 1.3~mm point spread function (psf). The combined Band 6+7 image at 1.0~mm and its corresponding spectral index map are shown in Figs.~\ref{hltau}e, and f (see \S~\ref{Details} for details). Notably, while the ring structure remains clearly visible and much improved over the 0.87~mm data alone, the combined 1.0~mm image does not show some of the (potentially interesting) azimuthal changes in ring brightness seen in the individual images. These might be due to variations in dust opacity, but they could also just be the result of differences in the uv-coverage at the two wavelengths. Thus, we strongly caution against over-interpretation of subtle features unless verified by detailed modeling or future even higher fidelity observations. 

The continuum properties of the \HLt\ disk are given in Table~\ref{Cont}. A wide range of flux densities is available in the literature. Considering recent interferometric observations, 2.7~mm flux densities range from $94.1\pm 0.9$ to $120\pm 4$~mJy and 1.3~mm flux densities range from $700\pm 10$ to $818\pm 10.8$ mJy \citep{Kwon2011,Guilloteau2011}. Using the more precise  2.7~mm measurement, and scaling by a spectral index $\alpha = 3$, there is good agreement with the 2.9~mm ALMA measurement ($74.3\pm 0.4$~mJy). At 1.3~mm, past observations bracket the total flux measured by ALMA ($744.1\pm 1.5$~mJy). At 0.87~mm, only one interferometric measurement is available: $1300\pm 300$~mJy \citep[SMA with $2\farcs1\times 1\farcs0$ resolution][]{Lumbreras2014}, but this is $60\%$ smaller than the ALMA measurement of $2140.8\pm 3.7$~mJy. In contrast, previous bolometric JCMT measurements at 0.86~mm wavelength have obtained $2360\pm 90$~mJy \citep{Andrews2005}, in much closer agreement with the ALMA data, especially considering the difference in angular resolution. This latter comparison suggests that 
only $\sim 10\%$ of the emission is resolved out by the ALMA data. 

Using the ALMA integrated flux densities from Table~\ref{Cont}, the average spectral index is $\alpha = 2.77\pm 0.13$, with no significant curvature evident from 2.9 to 0.87~mm, despite the expectation that the emission would be more optically thin at the longer wavelength. This result is likely a reflection of the fact that the central $\sim 10$~AU is both significantly brighter than the surrounding disk and is possibly optically thick even at 2.9~mm. Additionally, there may be weak free-free contamination in the central region at 2.9~mm \citep[see for example][]{Wilner1996,Carrasco2009}. For comparison, \citet{Andrews2005} find $\alpha=2.53\pm0.13$ based on a compilation of primarily bolometer measurements between 1.3 and 0.35~mm, suggesting the average spectral index does eventually flatten at shorter wavelengths. 

The \HLt\ disk is viewed on the sky inclined with respect to our line of sight (defined as $i=0\arcdeg$ for face-on), and rotated by an amount defined by the position angle (P.A.; measured from North to East). In order to constrain these parameters we first followed the standard practice of fitting a Gaussian to the visibility data.  Using baselines $< 1000$~m ($\sim 0\farcs2$), we obtain $i = 46.2\pm 0.2\arcdeg$ and P.A.$=138.2\pm 0.2\arcdeg$ in good agreement with past {\em uv-plane} estimates \citep{Kwon2011,Guilloteau2011}. However, the fantastic fidelity and angular resolution afforded by the ALMA images allow us to go further, and explore the properties of the rings independently in the {\em image-plane}.

\subsubsection{Spatially Resolved Disk Geometry and Spectral Index}
\label{geom}

\begin{figure}
\centering
\includegraphics[width=0.45\textwidth]{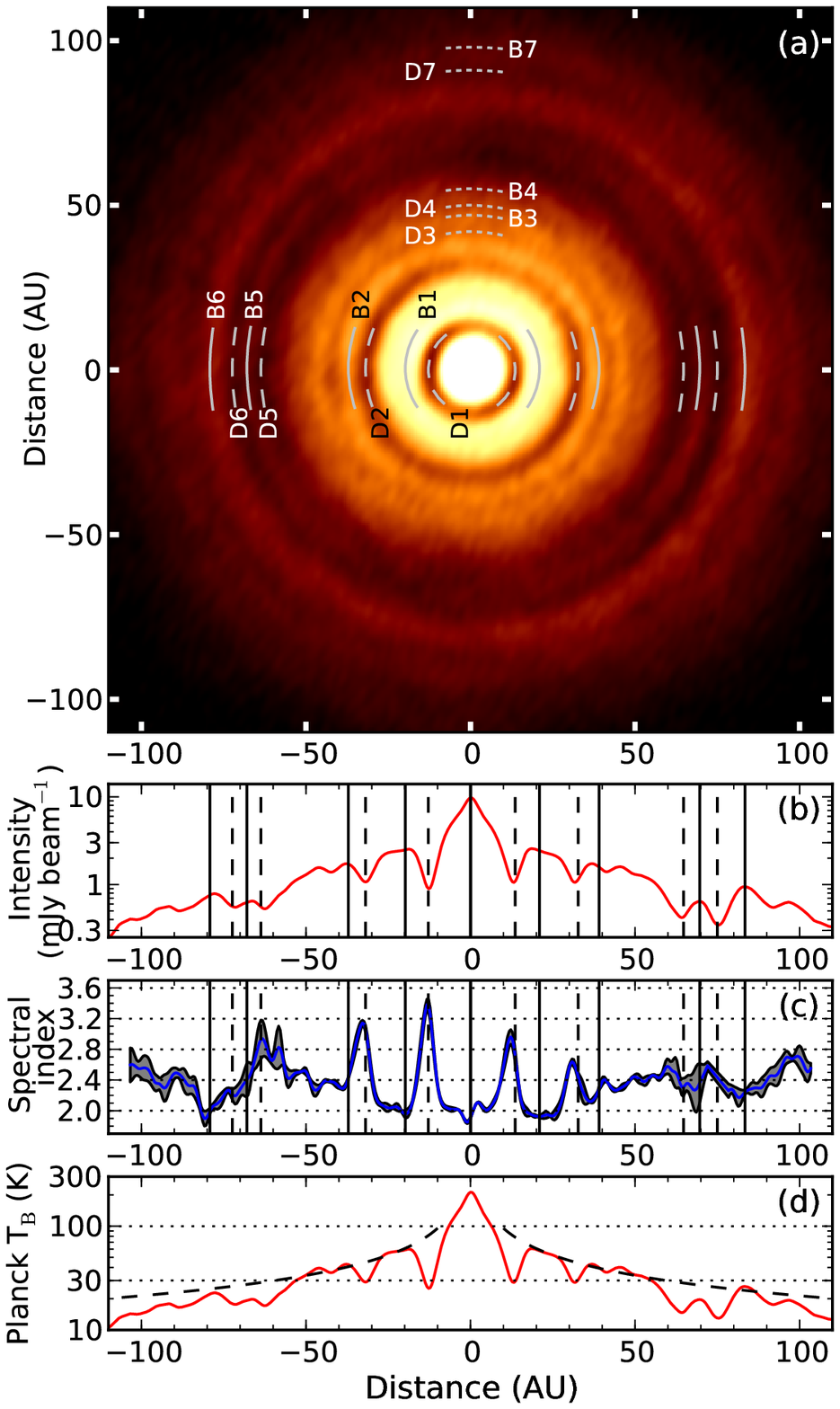}
\caption{Panel (a) shows the deprojected 1.0~mm B6+B7 image of
  \HLt\ (see \S3.1.3); the angular resolution is $38.6\times 19.3$~mas
  (PA.~$-20.7\arcdeg$).  Rings for which a full range of ellipse
  parameters could be fit are labeled horizontally (solid and dashed
  lines), while the less distinct rings are labeled vertically (dotted
  lines).  Panels (b) and (c) show cross-cuts at PA$=138\arcdeg$
  through the continuum peak of the 1.0~mm continuum and spectral
  index images shown in Fig.~\ref{hltau}e,f (positive values of
  distance correspond to the SE portion of Fig.~\ref{hltau}e). In
  panel (c) the grey region delineates the statistical $1\sigma$
  spectral index uncertainty, it does not account for the absolute
  flux uncertainty. For the fully fitted rings, panels (b) and (c)
  show dashed lines for the locations of dark rings and solid lines
  for bright rings. The fitted offsets from the continuum peak
  (Table~2) have been taken into account. Panel (d) shows the same
  cross-cut as panel (b) but on a Planck brightness temperature
  ($T_B$) scale. Panels (b) and (d) are shown on a log scale on the
  y-axis.  The dashed curve on panel (d) shows a representative power
  law for $T_B$ as a function of radius with an exponent of -0.65
  extending from the B1 peak, it is not a fit.}
\label{deproj}
\end{figure}

\begin{deluxetable}{lccccc}[t]
\tabletypesize{\footnotesize}
\tablewidth{0pc}
\tablecaption{Properties of Bright and Dark Rings\tablenotemark{a}}
\tablecolumns{6}
\tablehead{
\colhead{Ring} &  \multicolumn{2}{c}{Position Offset\tablenotemark{b}} & \colhead{Semi-major} & \colhead{Inclination} & \colhead{PA}\\   
\colhead{Name} &  \colhead{RA.(mas)} & \colhead{Dec.(mas)} & \colhead{(AU)} & \colhead{($\arcdeg$)} &  \colhead{($\arcdeg$)}
}
\startdata
D1  & $-2.0\pm0.9$ & $ -1.9\pm 0.9$ & $13.2 \pm 0.2$ & $44.9\pm1.1$ & $137.2\pm1.8$ \\
B1  & $-0.7\pm0.7$ & $ -4.9\pm 0.7$ & $20.4 \pm 0.1$ & $46.0\pm0.5$ & $138.2\pm0.9$ \\
D2  & $-0.9\pm0.5$ & $ -3.0\pm 0.6$ & $32.3 \pm 0.1$ & $45.5\pm0.3$ & $138.1\pm0.5$ \\
B2  & $-1.6\pm0.5$ & $ -7.9\pm 0.6$ & $38.1 \pm 0.1$ & $46.5\pm0.2$ & $137.3\pm0.4$ \\
D3  & & & $\sim 42$ & & \\
B3  & & & $\sim 47$ & & \\
D4  & & & $\sim 50$ & & \\
B4  & & & $\sim 55$ & & \\
D5  & $-1.8\pm0.4$ & $  3.4\pm 0.5$ & $64.2 \pm 0.1$ & $45.8\pm0.1$ & $138.9\pm0.2$ \\
B5  & $-0.8\pm0.5$ & $  7.5\pm 0.6$ & $68.8 \pm 0.1$ & $46.6\pm0.1$ & $138.8\pm0.2$\\
D6  & $-4.7\pm0.5$ & $  8.3\pm 0.5$ & $73.7 \pm 0.1$ & $47.9\pm0.1$ & $137.0\pm0.2$ \\
B6  & $-8.6\pm0.4$ & $ 12.3\pm 0.4$ & $81.3 \pm 0.1$ & $46.8\pm0.1$ & $137.9\pm0.1$ \\
D7  & & & $\sim 91$ & & \\
B7  & & & $\sim 97$ & & 
\enddata
\tablenotetext{a} {Rings with a complete set of parameters were calculated assuming circular orbits that are tilted by the inclination, and rotated in the sky by the P.A. using the method described in \S3.1.3. Rings with only a Semi-major axis entry were estimated directly from the image and have an uncertainty of order 1~AU.}
\tablenotetext{b} {Offset positions are with respect to the fitted 1.0~mm peak position in Table~\ref{Cont}.}
\label{rings}
\end{deluxetable}

Upon visual inspection, one can plausibly identify seven pairs of ``bright" and ``dark'' rings in the 1.0~mm image, we label these rings B1..B7 and D1...D7, respectively. 
As a first step, we assumed that the rings trace circular orbits around a common center position defined by the 1.0~mm continuum peak position (see Table~\ref{Cont}), and having the $i$ and P.A. derived in \S3.1.2. Then using a cross-cut along the major axis of the image, we determine approximate semi-major axes for the fourteen rings. 
%We refer to this combination of ring parameters as the ``zeroth-order ellipses'' below. 
After overlaying these zeroth-order ellipses on the image, it was apparent that while approximately correct, these are not a good fit to the rings in detail. To refine the parameters, we defined a discrete set of points along each zeroth-order ellipse, at a Nyquist sampled interval with respect to the synthesized beam.  Then the position of each point was moved to the nearest local radial maximum (or minimum for dark rings). To avoid regions where the rings become less distinct, points were discarded if they moved outside the nominal width of the individual rings (5 to 8~AU). Eight rings retained $>55\%$ of the points, to which we subsequently fit an ellipse, including its center position, using a Markov Chain Monte Carlo \citep[MCMC;][]{Foreman2013}. The results are listed in Table~\ref{rings}, with the full range of parameters given for the eight most distinct rings, and just the semi-major axis for the others. It seems likely that the ``gap'', ``enhancement'', and ``clump'' observed in VLA 1.3 and 0.7~cm images \citep{Carrasco2009,Greaves2008} at $\sim 10$, 20, and 55~AU along the major axis of the disk correspond to the D1, B1 and the combined emission from the  B2 to B4 rings, respectively.

The weighted average of the best-fit inclination and position angle for the eight fitted rings yields $i = 46.72\arcdeg\pm0.05\arcdeg$ and P.A.$=138.02\arcdeg\pm0.07\arcdeg$, consistent with the constraints found for the average disk geometry over large scales. However, the best-fit ellipses have their center's offset with respect to the peak of the 1.0~mm emission, as can be seen in the equatorial offsets reported in Table~\ref{rings}. These offsets are statistically significant for all but the innermost ring (D1). Interestingly, the magnitude of the position offset increases with orbital distance from the center.

Using the weighted average inclination and P.A., we have deprojected the combined 1.0~mm visibility data into a circularly-symmetric, face-on equivalent view (see Figure~\ref{deproj}a).  We have also extracted cross-cuts at an angle of $138\arcdeg$ from both the 1.0~mm continuum image and the spectral index map shown in Figs.~\ref{hltau}e,f. These cross-cuts are shown in Figs.~\ref{deproj}b,c. The variation in intensity between the bright and dark rings is readily apparent. Considering only the fully characterized rings, the largest average intensity contrast is between the first pair with D1 being $46\%$ less bright than B1, and the smallest contrast is between the 5th pair with D5 being $15\%$ less bright than B5. Such a drop in intensity could be due to a reduction in dust temperature, column density, or grain emissivity, or a combination thereof. This figure also demonstrates the general trend of having very high S/N on the estimate of $\alpha$ in the (brighter) inner parts of the disk with increasing uncertainty as the disk intensity decreases. Interestingly, the minimum $\alpha$ does not occur precisely at the continuum peak position, but is instead offset by $\sim 1.5$~AU ($\sim10.7$~mas) along the major axis to the SE. The origin of this offset is unknown.
% but it does not appear to be an artifact.

As shown in Figs.~\ref{deproj}b,c, each dark ring corresponds to a
local maximum in the spectral index, while each bright ring
corresponds to a local minimum. We do not find a gradual decrement of
the spectral index with radius reported in other lower mass
protoplanetary systems, albeit with reduced angular resolution
\citep[e.g. ][]{Guilloteau2011,Perez2012}. The central peak, the B1
ring, and the B6 ring, show a spectral index $\alpha\sim2$ indicative
of optically thick dust emission (within $3\sigma$). The observed
$T_B$ (see Fig.~\ref{deproj}d) provides a strict lower limit on the
physical temperature, at a given angular resolution. In the limit that
the emission is optically thick, $T_B$ provides a measure of the
physical temperature of the material where $\tau\approx 1$. The $T_B$
for the optically thick continuum peak, and azimuthally averaged
values for the optically thick B1 and B6 rings are 212.4$\pm 0.8$~K,
$59\pm 3$~K, and $24\pm 2$~K (corresponding intensities are $9.79\pm
0.04$, $2.48\pm 0.12$, and $0.85\pm 0.06$ \mjb\/), at radii of $\sim
0$, 20, and 81 AU, respectively.  As shown in Fig.~\ref{deproj}d, the
observed radial decrease in $T_B$ for all the bright rings can be
roughly characterized by a power law with an exponent of $\sim-0.65$.
For comparison, \citet{Guilloteau2011} found $T_B=25$~K at 55~AU from
$\sim 1\arcsec$ resolution PdBI data in good agreement with that
predicted from the ALMA $T_B$ analysis. However, the observed $T_B$
power-law should not be mistaken for a model of the physical dust
temperature, because much of the disk does not have a spectral index
consistent with optically thick emission.

The standard model used for circumstellar disks consists of a surface
layer of grains directly heated by stellar photons which subsequently
illuminates and heats the dust in the lower layers and the midplane
\citep{Chiang97}.  While the surface layer is optically thick to
stellar photons, it is optically thin at millimeter wavelengths. Thus,
the brightness temperature observed by ALMA in the optically thick
regions should correspond to the physical temperature at the $\tau
\approx 1$ surface, which will be somewhere between the surface
temperature ($T_s$) and the midplane temperature ($T_m$).  The
observed $T_B$ in the central peak and the B1 ring do indeed lie
between the values of $T_s$ and $T_m$ at those radii predicted from
the \citet{Kwon2011} ``thick-disk" model that best fits the full
HL~Tau SED.  Furthermore, the $T_B$ of all of the dark rings falls
below $T_m$, consistent with the dark rings being optically thin.  The
$T_B$ in the B6 ring is somewhat below the $T_m$ prediction of 33~K at
81~AU (despite having $\alpha\sim 2$). One possible interpretation for
this behavior is that B6 is not optically thick but instead has very
low $\beta$. However, we note that the \citet{Kwon2011} "thin-disk"
model, which provides the best match to the previous millimeter data alone (due
to dust settling to the mid-plane), predicts $T_m$ values that fall
below the observed $T_B$ (already a lower limit on the physical
temperature) in all of the dark rings. Clearly, more sophisticated
modeling of HL~Tau's physical dust temperature distribution that
includes the radial intensity and spectral index variations, as well
as radial changes in the scale-height will be fruitful.

As shown in Fig.~\ref{deproj}b,d, the dark rings are not completely devoid of emission, and the spectral index at their location differs from the optically thick expectation of $\alpha=2$, with larger $\alpha$ values ranging from $\sim2.3$ to $3.0$. This observation rules out the possibility that the reduction in emission is due {\em solely} to radial temperature variations.  Instead, these regions are likely optically thin, consistent with their lower observed $T_B$.  An estimate for the value of the dust opacity spectral index $\beta$ can be derived in the optically thin limit for warm dust where $\beta=\alpha-2 $ \citep{Beckwith1991}. We estimate 
$\beta\sim0.8$ for D1, 
$\beta\sim0.7$ for D2, 
$\beta\sim0.6$ for D5 and 
$\beta\sim0.3$ for D6, consistent with some amount of grain growth and evolution inside these dark rings. However, further physical modeling of the disk dust continuum emission is needed to confirm these estimates. Finally, we caution that observations which cannot resolve the \HLt\/ morphology will combine emission from optically thick and thin regions, driving any derived values of dust opacity spectral index to be lower than reality.

\begin{figure*}
\centering
\includegraphics[width=0.9\textwidth]{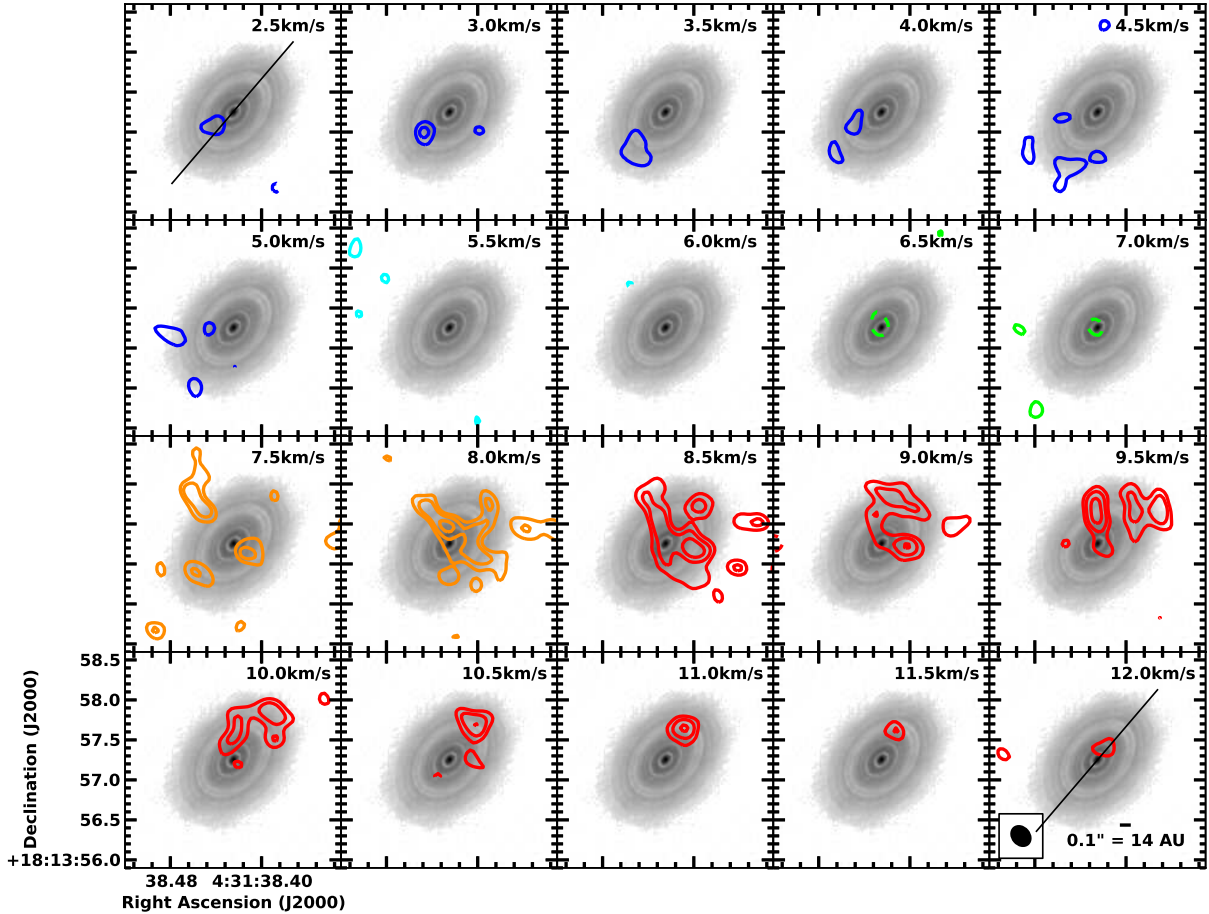}
\caption{Contours show channel maps of the \hcop\/ emission with $\sim 0\farcs25$ angular resolution and 0.5~\kms\/ velocity resolution overlaid on the 1.0~mm B6+B7 continuum image in greyscale. The LSRK velocity of the emission is shown in the upper right corner of each panel.  The contour levels are at 1.9($1\sigma$)$\times$ (-3.5, 3.5, 5.0, 7.0) \mjb\/ and the color indicates the change from blue- to red-shifted emission. Negative contours are shown dashed. The cross-cut position angle of $+138\arcdeg$ is shown on the 2.5 and 12.0 \kms\/ panels and the \hcop\/ synthesized beam is shown in the lower left corner of the last panel.\\
\\
}
\label{channel}
\end{figure*}

\begin{figure}
\centering
\includegraphics[width=0.4\textwidth]{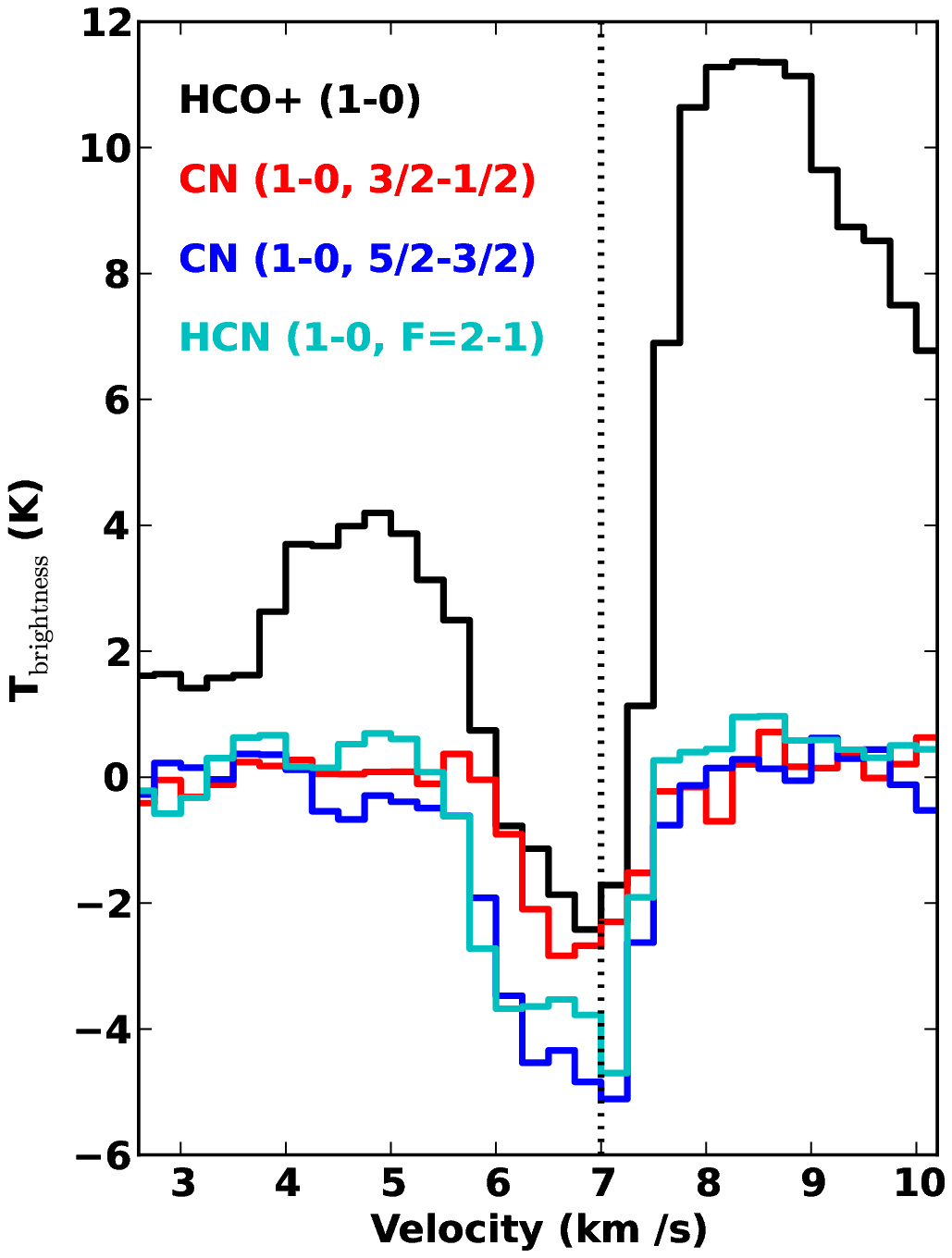}
\caption{Continuum-subtracted spectra with angular resolution $1\farcs$1 from the indicated transitions toward the \HLt\ continuum peak. The adopted LSRK systemic velocity of 7~\kms\/ is indicated by the vertical dotted line. For comparison the 2.9~mm peak continuum surface brightness at $1\farcs1$ resolution is $5.75\pm 0.01$~K.}
\label{abpv}
\end{figure}

\subsubsection{Compact Spectral Lines}

At $1\farcs1$ resolution, the \hcop\/ emission shows a morphology similar to the \co\/ shown in Fig.~\ref{CO}a, albeit somewhat less extended. Interestingly, it also shows a barely resolved velocity gradient across the \HLt\ disk itself, though it is confused with the surrounding outflow gas, especially to the NE of the disk. Fortunately, at $0\farcs25$ resolution most of the outflow emission is resolved out, making it possible to spatially resolve the morphology of the \HLt\ molecular gas disk for the first time. The $0\farcs25$ resolution \hcop\/ channel maps shown in Figure~\ref{channel} reveal: (1) A notable deficit of blue vs. red-shifted emission; (2) \hcop\/ absorption near the systemic velocity 6.5-7.0 \kms\/; (3) A roughly Keplerian velocity distribution with a detectable radial velocity range of 2.0 to 12.0 \kms\/; (4) Comparable gas and continuum disk sizes (at least at the current sensitivity level). To further explore points (1) and (2), Figure~\ref{abpv} shows spectra from the $1\farcs1$ resolution cubes toward the continuum peak. All of the lines show absorption  at 6.0 to 7.0 \kms\/, indeed CN and HCN (1-0) are only detected in absorption. Additionally, the absorption is non-Gaussian in shape, instead showing a gradual increase in absorption on the blue-shifted side compared to a steeper rise on the red-shifted side. As shown in Fig.~\ref{CO}a, the blue-shifted outflow emerges from the NE of \HLt\ and, based on the continuum disk orientation, propagates toward us at $i \approx 47\arcdeg$. Self-absorption by this outflowing gas is likely responsible for both the non-Gaussian line shape and the deficit of observable blue-shifted disk emission.  The deepest absorption for the CN and HCN transitions occurs at a LSRK velocity of $7.0\pm0.2$ \kms\/ which we take to be the rest velocity of the system. 

Under the assumption of circular Keplerian motion, and noting that the velocity extrema of \hcop\/ emission occurs at $\Delta V\pm 5$~\kms\/ (see Fig.~\ref{channel}) from the systemic velocity (7.0~\kms\/) at a radius of $\sim 25$~AU, we find that with $i=47\arcdeg$, the enclosed mass is $\sim 1.3$~\msun\/. This value is near the high end of the range previously reported for HL~Tau (see \S1). However, it is clear that even on the less absorbed red-shifted side, the velocity pattern is not so simple, and may for example have a contribution from infall \citep[see for example][]{Gomez2000}. Future detailed radiative transfer analysis coupled with a physical model will be required to reproduce the complex \hcop\/ absorption and emission toward the \HLt\/ disk in order to obtain a more accurate kinematic stellar mass.

\subsection{XZ~Tau 2.9~mm Continuum}

At 2.9~mm we resolve the known multiple system XZ~Tau into two components, A and B \citep[Fig.~\ref{CO}b, Table~\ref{Cont}, also see][]{Forgan2014,Carrasco2009} separated by 273$\pm1$~mas at a position angle of 128.7$\pm0.5\arcdeg$. This separation is $8\%$ smaller than predicted by \citet[][]{Forgan2014}  for a circular, face-on orbit (296$\pm1$~mas), particularly in R.A. This is likely an indication that the orbit is not face-on, but this will require future observations to confirm and quantify. Like \citet{Forgan2014}, we find no evidence for component ``C" (a putative third star) reported by \citet{Carrasco2009} at 7~mm. Using the 2.9~mm flux densities from Table~\ref{Cont} and the JVLA 7~mm flux densities from \citet{Forgan2014}, we find spectral indices of $+1.8\pm 0.5$ for both XZ~Tau A and B, suggesting both have a free-free component in addition to dust emission \citep[see also][]{Carrasco2009}.

\subsection{\Lk\/ 2.9~mm Continuum} 

At 2.9~mm, we have resolved the \Lk\/ disk for the first time in millimeter continuum to a size of only 
$21\pm 1$~AU with an inclination angle of $56\pm 2\arcdeg$ at a PA.$=170\pm 3\arcdeg$ (see Fig~\ref{CO}c, Table~\ref{Cont}). \citet{Schaefer2009} used PdBI to observe \cotwoone\/ with a resolution of $1\farcs38\times 0\farcs83$ toward this source, obtaining an inclination of $28\pm 9\arcdeg$, $V_{100}$sin$i$=1.35$\pm 0.04$~\kms\/ (the radial velocity at a radius of 100 AU), and a dynamical mass of 0.5-2 M$_{\sun}$. These authors note that the inclination angle is not well-constrained by the PdBI data. If we assume instead that the molecular gas has the same inclination as the continuum the dynamical mass is significantly smaller, $0.3\pm 0.1$~M$_{\sun}$. The ALMA 2.9~mm integrated flux density measured for \Lk\/ ($3.7\pm 0.1$~mJy) is in reasonable agreement with that measured by \citet{Schaefer2009} at 2.7~mm: $4.0\pm 0.6$~mJy.

\section{Discussion}

What is the nature of the remarkable pattern of bright and dark rings in the continuum emission from the \HLt\ protoplanetary disk? Here we highlight four key observational findings. First, the correlated radial structure of higher spectral index and lower brightness temperature in the dark rings, relative to the bright rings, demonstrates that the optical depth must be lower in the dark rings.  Second, the fact that the centers of nearly all of the rings are significantly offset from the dust continuum peak provides compelling evidence that the rings are not circular as assumed, but rather have some eccentricity with \HLt\/ at one focus.  Third, the magnitude of the offset increases with ring radius, which is tantalizingly congruent with the observed increase in orbital eccentricity for exoplanets at large orbital radii \citep[][]{Butler2006,Shen2008}. Fourth, several of these rings appear to be in resonances with each other.  Assuming Keplerian orbits and neglecting the mass between the rings, the first four dark rings have orbital periods in the proportion D1:D2:D3:D4 = 1:4:6:8.  There also appear to be resonances between bright and dark rings, with D2 in a 2:1 resonance with B1 and in a 1:4 resonance with B6\footnote{Other bright rings are close but not exactly at the following integer resonance ratios: D2:B2:B3 are near to a 3:4:5 resonance, while D2:B5 are close to a 1:3 resonance.}. These mean motion resonances (MMRs) are calculated to be within the confidence intervals obtained on the semi-major axes. Given the precise constraints obtained on the orbital radius of the bright and dark rings, the n:m resonances found here have uncertainties between 0.01 and 0.09, making these resonance ratios unlikely to arise from random chance. Although multi-planetary systems show a diverse architecture\footnote{Note that HR8799 is the only currently known multi-exoplanet system spanning a comparable range of semi-major axes as the gaps in the \HLt\ disk \citep[see][and references therein]{Gozdziewski14}. The planets in other multiple systems orbit within $\sim5$~AU of the host star.} and many pairs of exoplanets are far from being in MMR \citep{Lissauer11}, a growing number of stable systems do exist in or near MMR \citep{Zhang2014}.  Gravitational interactions between planets and their parent disk can lead to resonant and near-resonant systems \citep[see][and references therein]{Baruteau2013}.  Collectively, these four independent lines of evidence suggest that the observed dark rings are low column density gaps in the disk material arising from the process of planet formation.

\section{Conclusions}
In this paper, we present multiwavelength ALMA images of the \HLt\ protoplanetary disk at resolutions as fine as a few AU. We derive precise measurements of the disk inclination and position angle, as well as the proper motion of the central protostar.  We also show the first molecular line observations to spatially resolve the disk kinematics.  Under the assumption of Keplerian rotation, the mass of the central star is $M_*\sim1.3$~M$_{\sun}$, but further refinement will require detailed radiative transfer calculations. The remarkable pattern of bright and dark circumstellar rings in the continuum images exhibits a corresponding structure in the spectral index image, revealing that the dust in the central core and several of the bright rings are optically thick, while the dust in the dark rings -- which are not completely empty -- shows evidence of grain growth.  Several characteristics of these rings, including an increase in eccentricity with radius and numerous resonances, suggest that the dark rings are gaps arising from the process of planet formation.  Modeling of the disk temperature and density structure in both gas and dust, coupled with planet formation processes will be crucial to the goal of reproducing \HLt\/'s morphology from theory \citep[e.g.,][]{Dong2014}. These transformational ALMA observations of \HLt\/ herald a new era in the study of protoplanetary disks that promises to unearth the architecture of extrasolar multi-planetary systems during their epoch of formation, and thereby illuminate the origin of our own Solar System.

\acknowledgements This paper makes use of the following ALMA data: \dataset{ADS/JAO.ALMA\#2011.0.00015.SV}.
ALMA is a partnership of ESO (representing its member states), NSF (USA) and NINS (Japan), together with NRC (Canada), NSC and ASIAA (Taiwan), and KASI (Republic of Korea), in cooperation with the Republic of Chile. The Joint ALMA Observatory is operated by ESO, AUI/NRAO and NAOJ. The National Radio Astronomy Observatory is a facility of the National Science Foundation operated under cooperative agreement by Associated Universities, Inc. This research made use of Astropy, a community-developed core Python package for Astronomy \citep{Astropy2013}, as well as the VizieR catalogue access tool, CDS, Strasbourg, France.. This paper also makes use of observations made with the NASA/ESA Hubble Space Telescope, and obtained from the Hubble Legacy Archive, which is a collaboration between the Space Telescope Science Institute (STScI/NASA), the Space Telescope European Coordinating Facility (ST-ECF/ESA) and the Canadian Astronomy Data Centre (CADC/NRC/CSA).

{\it Facilities:}\facility{ALMA}.

\appendix

\section{Imaging Details}
\label{Details}

The data were calibrated in CASA~4.2.2 using the same procedure adopted for standard science operations, with two exceptions. An imprecise position was inadvertently used for the phase calibrator J0431+2037 employed for the 2.9 and 1.3~mm data.  The shift was small ($\sim 15$~mas), but noticeable when compared to the 0.87~mm data, for which a precise position for J0431+1731 was employed. The position of J0431+2037 was corrected to (J2000) 04$\h$31$\m$03.76137$\s$ $+$20$\arcdeg$37$\arcmin$34.2649$\arcsec$ before calibration, after which excellent position agreement for HL~Tau was achieved across all three Bands (see fitted positions in  Table~\ref{Cont}). Both 
final phase calibrator positions were taken from \url{http://astrogeo.org/vlbi/solutions/rfc\_2012b/rfc\_2012b\_cat.txt}. The small shifts in position due to proper motion over the month long observing span have been ignored. Additionally, the final LBC antenna position corrections were applied before the data were calibrated \citep[see ][]{ALMA15a}. For each execution, the bandpass and absolute flux calibrators were dynamically chosen by the ALMA online system and alternated between the frequently monitored quasars J0423-0120 and J0510+1800, depending on the LST start time$^{22}$. Based on comparison of the resulting calibrated phase calibrator flux densities across the many executions for each band, we estimate that the absolute flux calibration is good to within $5\%$ for the final combined data at each wavelength.
 % (uncertainties for any individual execution are significantly larger). 

The continuum images for all three bands were made using multi-frequency synthesis, and the visibility weighting employed was 
mid-way between natural and uniform (i.e., CASA {\em clean} parameter {\em robust=0.0}).
Individually, the 1.3~mm and 0.87~mm data have a fractional bandwidth small enough  ($< 10\%$) to ignore spectral index effects in the imaging. In contrast, the range of continuum spws available for the 2.9~mm image (90.8 to 102.9 GHz), if all combined, would require accounting for at least a linear spectral index slope across the band, given the expected range of dust spectral indices $\alpha =2-4$ ($S_{\nu}\propto {\nu}^{\alpha}$). However, inclusion of the spectral slope requires good signal-to-noise ratio (S/N) at both frequency extrema of the calculation \citep[][]{Rau2011}, a condition the 2.9~mm data did not fulfill since four of the five available continuum spws were clustered at $\sim 102$ GHz. Thus, the 2.9~mm continuum image was created excluding the 90.8~GHz spw and without accounting for the spectral index. To make an image with the highest possible combination of fidelity and angular resolution we also made a combined Band~6+7 (1.0~mm) image (fractional bandwidth $44\%$) accounting for spatial variations in the spectral index (CASA {\em clean} parameter {\em nterms=2}). In this case, the S/N is high at both extrema and the resulting image and spectral index map are reliable (the derived $\alpha$ have a statistical S/N of up to several hundred). Additionally, we used the {\em multiscale} imaging option \citep[][]{Cornwell2008} for all three bands with scales of 0, 5, and 15, approximately corresponding to 0, 1, and 3$\times$ the synthesized beam. Using these imaging parameters, the combined continuum data at each band was iteratively self-calibrated.  
The continuum peak position in all bands changed by $< 0.8$~mas following self-calibration. The final synthesized beams and rms noise levels are 
given in Table~\ref{Cont}.

%\footnote{Due to the non-optimal antenna configuration available for the LBC that concentrated baselines at lengths $< 200$~m and $> 500$~m, the choice of visibility weighting is an important one. For example, though it yields the best sensitivity in principle, natural weighting ({\em robust=2.0}) results in a very non-Gaussian point-spread function (psf; i.e., dirty beam) with a significant $\sim 1\arcsec$ plateau underlying the compact peak that corresponds to the fitted clean beam, which leads to poor image reconstruction.}

Even with ALMA's impressive collecting area, the surface brightness sensitivity at very high angular resolution is limited for detecting narrow thermal lines. To improve the surface brightness sensitivity, the data for the four Band~3 line transitions were tapered to a resolution of $\sim1\farcs1$ (and later convolved to exactly this value) and imaged with {\em robust=0.5}, and a velocity width of 0.25~\kms\/. The \co\/ and \hcop\/ lines both show significant extended emission at this resolution and were cleaned with the same multiscale parameters as the continuum (an additional scale of 45 was used for CO). The \cn\/ and \hcn\/ lines are only detected in absorption toward the continuum peak. The resulting  rms noise levels for the $1\farcs1$ resolution HCN (1-0), \hcop\/, CN (1-0), and \co\/ cubes are
2.5, 2.5, 5.0, and 9.0~\mjb\/, respectively (the $1\farcs1$ taper reduces the effective number of antennas to $\approx 20$). Additionally, an \hcop\/ line cube was made at a taper corresponding to an angular resolution of $\sim 0\farcs25$ ($0\farcs28\times 0\farcs22$, PA$=+40.5\arcdeg$) with {\em robust=0.0}, and 0.25~\kms\/  channels in order to explore the compact disk emission; the rms noise in this cube is 2.4~\mjb\/. At this resolution \co\/ still shows significant confusing contamination from outflow emission as well as copious self-absorption, while the \cn\/ and \hcn\/ lines are only marginally detected.  We note that the continuum self-calibration was not applied to the line data, as tests demonstrated that this did not improve the S/N.

%\footnote{For example, at 89~GHz with 0.25 \kms\/ channels, 30 antennas, 3.5 hours on-source, and an angular resolution of 70~mas, the expected spectral rms noise per channel is 2.6~\mjb\/ but only 83~K in surface brightness sensitivity.}

\end{document}